\documentclass[conference,a4paper]{APSIPA2018}
\usepackage{multirow}
\usepackage[dvips]{graphicx}
\usepackage{amsmath}
\usepackage[psamsfonts]{amssymb}
\usepackage{amsxtra}
\usepackage{threeparttable}
\usepackage{subfigure}
\usepackage{cite}
\usepackage{url}

\begin{document}

\title{Compressible and Learnable Encryption for Untrusted Cloud Environments}

\author{%
\authorblockN{%
Hitoshi Kiya
}
\authorblockA{%
Tokyo Metropolitan University, Tokyo, Japan\\
E-mail: kiya@tmu.ac.jp
}
}

\maketitle
\thispagestyle{empty}
With the wide/rapid spread of distributed systems for information processing, such as cloud computing and social networking, not only transmission but also processing is done on the internet. Therefore, a lot of studies on secure, efficient and flexible communications have been reported. Moreover, huge training data sets are required for machine learning and deep learning algorithms to obtain high performance. However, it requires large cost to collect enough training data while maintaining people’s privacy. Nobody wants to include their personal data into datasets because providers can directly check the data. Full encryption with a state-of-the-art cipher (like RSA, or AES) is the most secure option for securing multimedia data. However, in cloud environments, data have to be computed/manipulated somewhere on the internet. Thus, many multimedia applications have been seeking a trade-off in security to enable other requirements, e.g., low processing demands, and processing and learning in the encrypted domain,

Accordingly, we first focus on compressible image encryption schemes, which have been proposed for encryption-then-compression (EtC) systems, although the traditional way for secure image transmission is to use a compression-then encryption (CtE) system. EtC systems allow us to close unencrypted images to network providers, because encrypted images can be directly compressed even when the images are multiply recompressed by providers. Next, we address the issue of learnable encryption. Cloud computing and machine learning are widely used in many fields. However, they have some serious issues for end users, such as unauthorized access, data leaks, and privacy compromise, due to unreliability of providers and some accidents. 

\section{Compressible Image Encryption, EtC System}
A block scrambling-based image encryption scheme has been proposed for EtC systems with the assumption of the JPEG standard as a compressible image encryption scheme [1], in which a user wants to transmit image $I$ securely to an audience, via a social networking Service (SNS) provider like Twitter or a cloud photo storage service (CPSS) such as Google Photos, as illustrated in Fig.1. 
\begin{figure}[b]
\centering
\includegraphics[width = 0.9\columnwidth]{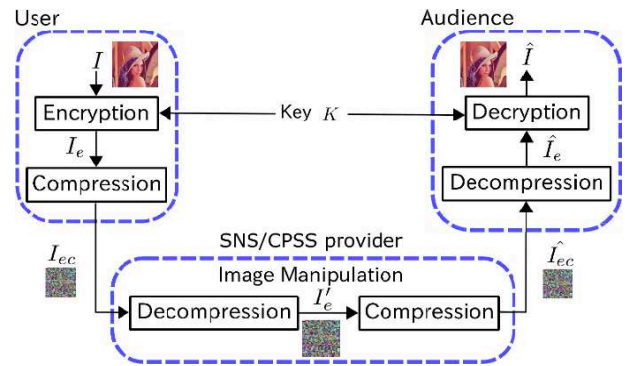}
\caption{EtC system}
\end{figure}
Because the user does not give the secret key $K$ to the provider, the privacy of image $I$ to be shared is under control of the user even when the provider recompresses image . Therefore, the user is able to control image privacy for his own demand. However, in CtE systems, the user has to disclose unencrypted images to recompress them. 

Figure 2 illustrates the procedure of the block scrambling-based encryption with the block size of 16 $\times$ 16, which consists of four encryption steps. An example of an encrypted image is shown in Fig.3(b); Fig.3(a) is the original one. This Image encryption scheme has been extended as a grayscale-based image encryption one to enhance the security of EtC systems [2]. An example of an encrypted image is shown in Fig. 3 (c), in which the block size is smaller, the number of blocks is larger, and the encrypted image includes less color information, than Fig.3 (b). Images encrypted using these schemes have the following properties.
\begin{description}
  \item[(a)] The compression efficiency for encrypted images is almost the same as that for the original ones under the use of JPEG compression.
  \item[(b)] Robustness against various attacks has been demonstrated.
\end{description}
Figure 4 shows the rate-distortion (RD) curves of JPEG compressed images without any encryption and with the block scrambling-based encryption, where the average bitrate and PSNR values of 20 images are plotted, after decrypting the images. The encrypted images were found to not be affected by JPEG compression. 
\begin{figure}[b]
\centering
\includegraphics[width = 0.9\columnwidth]{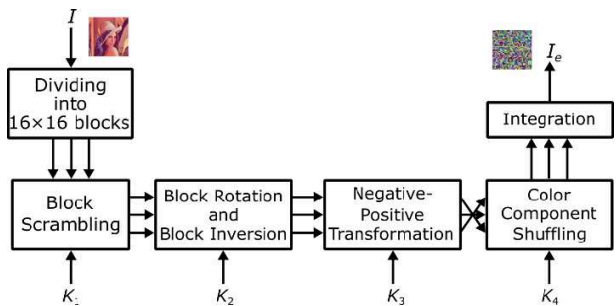}
\caption{Block scrambling-based image encryption}
\end{figure}
\begin{figure}[t]
	\centering
	\subfigure[Original image]{
		\label{subfig:re1-a}
		\includegraphics[width = 0.3\columnwidth]{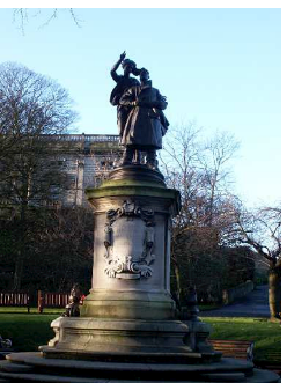}}
	\subfigure[Encrypted image]{
		\label{subfig:re1-b}
		\includegraphics[width = 0.3\columnwidth]{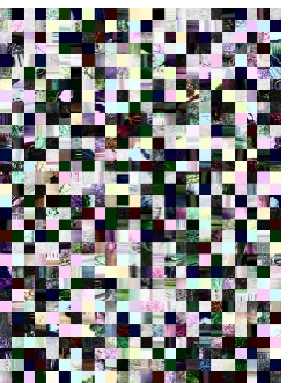}}
	\hspace{20mm}
	\subfigure[Grayscale-based encrypted image]{
		\label{subfig:re1-c}
		\includegraphics[width = 0.45\columnwidth]{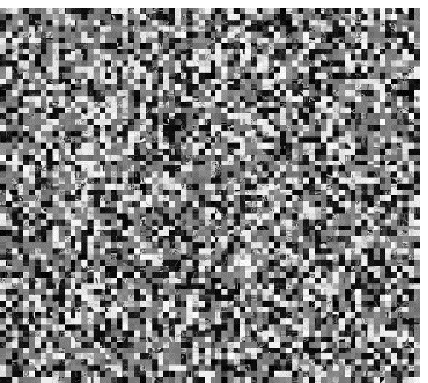}}
	\caption{Example of encrypted images}
	\label{fig:re_img}
\end{figure}
\begin{figure}[tbh]
\centering
\includegraphics[width = 0.9\columnwidth]{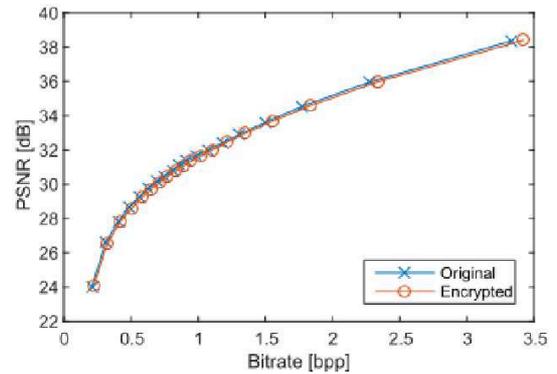}
\caption{RD curves of original images and encrypted ones}
\end{figure}

\section{Application to SNS and CPSS}
SNS providers like Twitter and Facebook, and CPSS providers like Google Photos are generally well known to manipulate images uploaded by users, and to support the JPEG standard, as one of the most widely used image compression standards. For this reason, images protected by almost all encryption schemes such as RSA are inapplicable to the providers, while images encrypted by the aforementioned schemes are applicable to most of the providers if encrypted images meet some conditions. 

In [3], EtC systems with the block-based image encryption have been applied to SNS providers. In one experiment, encrypted and non-encrypted JPEG images were uploaded to various SNS providers to determine the robustness of the EtC systems under various conditions. Moreover, the paper investigated how each SNS provider manipulates uploaded images, and the encryption schemes used in the EtC systems were evaluated in terms of the robustness against image manipulation by SNS providers. 

\section{Security Evaluation against Jigsaw Puzzle Solver Attacks}
Security analysis for EtC systems is needed, because the encryption schemes used in EtC systems do not have provable security. Therefore, safety has been evaluated first based on its key space assuming brute-force attacks, and the schemes have generally been shown to have enough key spaces for protecting against such attacks. However, each block in encrypted images has almost the same correlation as that of original images, which are needed to maintain a high compression performance. Jigsaw puzzle solvers, which utilize the correlation between pieces, have been actively studied in the area of computer vision, and they have succeeded in solving puzzles with a large number of pieces. We can regard the blocks of an encrypted image as pieces of a jigsaw puzzle. In [4]-[5], jigsaw puzzle solver attacks were discussed in addition to brute-force attacks, as a ciphertext-only attack (COA). Some solvers have been shown to be able to decrypt encrypted images even when the key space is large enough. However, assembling jigsaw puzzles becomes difficult under the following conditions.
\begin{description}
  \item[(a)] The number of blocks is high. 
  \item[(b)] The block size is small.
  \item[(c)] The encrypted images include JPEG distortion.
  \item[(d)] The images have no color information
\end{description}

Other attacking strategies such as known-plaintext attack (KPA) and chosen-plaintext attack (CPA) should be considered for the security. Block scrambling-based image encryption becomes robust against KPA through assigning a different key to each image for the encryption. In addition, the keys used for the encryption do not need to be disclosed because the encryption scheme is not public key cryptography. Therefore, the encryption can avoid the CPA unlike public key cryptography. 

Figure 5 illustrates examples of an encrypted image and the assemble images, where three measures ware used to evaluate the results: direct comparison $(D_C)$ , neighbor comparison $(N_C)$ , largest component $(L_C)$ . In the measures, $D_C, N_C, L_C \in \left[0,1\right]$ , a larger value means a higher compatibility. Encryption with four steps in Fig. 1 was shown to make assembling images more difficult, than encryption with one step, i.e. with only block scrambling. 
\begin{figure}[t]
	\centering
	\subfigure[Original image\newline  ~]{
		\label{subfig:re1-a}
		\includegraphics[width = 0.45\columnwidth]{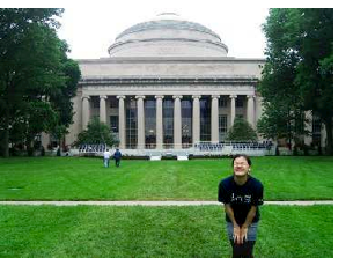}}
	\subfigure[Encrypted image \newline(4 steps encryption)]{
		\label{subfig:re1-b}
		\includegraphics[width = 0.45\columnwidth]{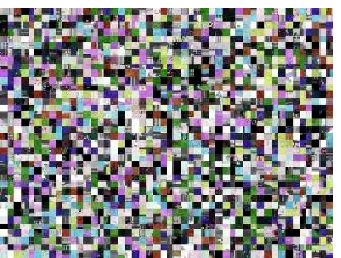}}
	\hspace{20mm}
	\subfigure[Assembled image \newline(1 step encryption)\newline($D_c = 0, N_c = 0.3, L_c = 0.1$)]{
		\label{subfig:re1-c}
		\includegraphics[width = 0.45\columnwidth]{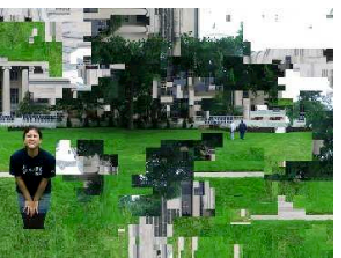}}
	\subfigure[Assembled image \newline(4 steps encryption)\newline($D_c = 0, N_c = 0, L_c = 0$)]{
		\label{subfig:re1-d}
		\includegraphics[width = 0.45\columnwidth]{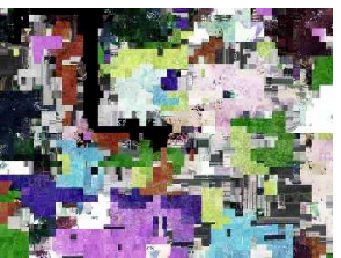}}
	\caption{Examples of encrypted images and assembled images}
	\label{fig:re_img}
\end{figure}

\section{Learnable Image Encryption}
Considerable efforts have been made in the fields of fully homomorphic encryption and multi-party computation [6]. However, these encryption schemes are still difficult to be applied to learning algorithms, although some attempts have been made to deep learning [7]-[8]. Moreover, the schemes require preparing algorithms specialized for computing encrypted data, and high computational complexity. In addition, the latest book on the subject [9] demonstrates the severity of the problem by providing a taxonomy of attacks and studies of adversarial learning. It also analyzes conventional attacks as well as the latest discovered weaknesses in deep learning systems.

Furthermore, privacy preserving computing schemes without homomorphic encryption and multi-party computation have also been considered for machine learning and deep learning [10]-[12]. They allow not only a light-weight computing cost, but also direct the computation of typical learning algorithms, without preparing any algorithms specialized for secure computing. In those methods, the block scrambling-based encryption in Fig.1 plays an important role as it does in EtC systems. Figure 6 illustrates the scenario of the privacy preserving computing. In the enrollment, client $i$, prepares training samples $g_i$, such as images, and a feature set ${\bf f}_{i,j}$, called a template, is extracted from the samples. Next the client creates a protected template set ${\hat {\bf f}}_{i,j}$ using a secret key $p_i$ and sends the set to a cloud server. The server stores it and implements learning with the protected templates for a machine learning algorithm. In the authentication, Client $i$ creates a protected template as a query and sends it to the server. The server carries out a classification problem with a learning model prepared in advance, and then it returns the result to Client $i$. Note that the cloud server has no secret keys and that the classification problem can be directly carried out using well- known algorithms.
\begin{figure}[tbh]
\centering
\includegraphics[width = 0.9\columnwidth]{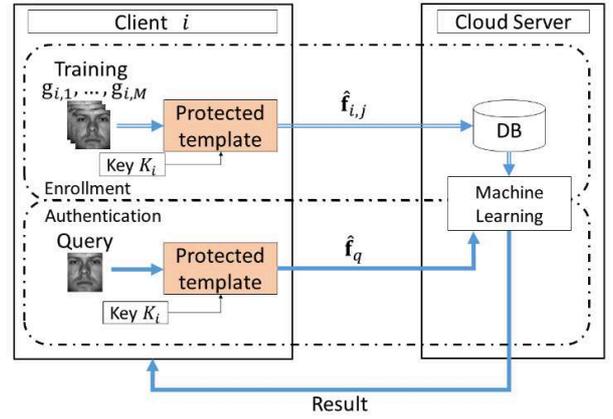}
\caption{Learnable encryption for machine learning}
\end{figure}

\section{Long-term Impact}
Communications, computing and data storage environments have been dramatically changing. A variety of defenses have to be considered for learning systems and various attack types in the new environments, although considerable efforts have been made in the field of information security and forensics.

For example, the EtC systems described in this article are applicable to only still images, so EtC systems for video data have not been developed yet, due to the difficulty in performing motion estimation algorithms in the encrypted domain. Moreover, most signal processing-friendly encryption schemes such as block scrambling-based ones have no provable security, but like our house keys, they are absolutely necessary for our lives. Therefore, new evaluation measures of the safety should be discussed, because encryption schemes without provable security have a lot of attractive features. Robustness against jigsaw puzzle solver attacks is one of the measures.

Machine learning and deep learning systems are very powerful tools in many fields. However, huge training data sets are required for the learning, and the data, such as surveillance data are generally sensitive. Therefore, privacy-preserving computing schemes are required to utilize various learning systems safely for our lives. Unfortunately, those computing schemes have not been sufficiently developed yet. In other words, a lot of research subjects still need to be conducted in this field. That is very fortunate for researchers.

\end{document}